\documentclass[sigconf]{acmart}

\usepackage{geometry}
\usepackage[ruled,linesnumbered,lined]{algorithm2e}
\usepackage{multirow}
\usepackage{url}
\usepackage{fontawesome}
\usepackage{caption}
\usepackage{amsmath}
\usepackage{etoolbox}
\usepackage{fancyhdr}
\usepackage{booktabs} 
\usepackage{enumerate}
\usepackage{color}
\usepackage{verbatimbox,xcolor,lipsum}
\usepackage{listings}
\usepackage{tcolorbox}
\usepackage{listings}
\usepackage{hyperref}
\usepackage{algorithmic}
\usepackage{graphicx}
\usepackage{textcomp}
\usepackage{xcolor}
\usepackage{tcolorbox}
\usepackage{color}
\usepackage{booktabs}
\usepackage{fontawesome}
\usepackage{url}
\usepackage{caption}
\usepackage{subcaption}
\definecolor{anti-flashwhite}{rgb}{0.95, 0.95, 0.96}
\definecolor{lightgray}{rgb}{0.83, 0.83, 0.83}

\AtBeginDocument{%
  \providecommand\BibTeX{{%
    \normalfont B\kern-0.5em{\scshape i\kern-0.25em b}\kern-0.8em\TeX}}}

\copyrightyear{2024}
\acmYear{2024}
\setcopyright{acmlicensed}
\acmConference[ICSE-Companion '24]{2024 IEEE/ACM 46th International Conference on Software Engineering: Companion Proceedings}{April 14--20, 2024}{Lisbon, Portugal}
\acmBooktitle{2024 IEEE/ACM 46th International Conference on Software Engineering: Companion Proceedings (ICSE-Companion '24), April 14--20, 2024, Lisbon, Portugal}
\acmDOI{10.1145/3639478.3643116}
\acmISBN{979-8-4007-0502-1/24/04}
\begin{document}

\title{xNose: A Test Smell Detector for C\#}

\author{Partha P. Paul, Md Tonoy Akanda, M. Raihan Ullah, Dipto Mondal, Nazia S. Chowdhury, and Fazle M. Tawsif}
\affiliation{%
  \institution{\{partha,raihan, tawsif\}-iict@sust.edu ; \{tonoy.sust, diptomondal007, nazia.nishat1971\}@gmail.com}
    \city{Shahjalal University of Science and Technology}
  \country{Bangladesh}
}

%
\title{xNose: A Test Smell Detector for C\#}



\newcommand{\tawsif}[1]{\textcolor{red}{(Tawsif: #1)}}

\begin{abstract}
Test smells, similar to code smells, can negatively impact both the test code and the production code being tested. While existing research has focused on identifying test smells in languages like Java, Scala, and Python, there is currently a lack of automated tools for detecting test smells in C\#, which has gained popularity in recent times. This paper aims to bridge this gap by extending the study of test smells to C\#, and developing a tool to identify test smells in this language and analyze their distribution across projects.

To start, we compiled a list of test smells from prior studies, selecting those that were language-independent and had equivalent features in the standard \textit{xUnit} framework for C\#. In total, we identified 16 distinct test smells. To facilitate this research, we built a tool called \textit{xNose}, which is an extension for Visual Studio capable of detecting specific test smells. Our evaluation of \textit{xNose} demonstrated a high level of accuracy, with an average precision score of 96.97\% and an average recall score of 96.03\%.

Furthermore, we conducted an empirical investigation to determine the prevalence of test smells in \textit{xUnit}-based C\# projects. This examination offers valuable insights into the frequency and distribution of test smells, providing a deeper understanding of their impact on C\# projects and test suites. The development of \textit{xNose} and our analysis of test smells in C\# code aim to support developers in maintaining code quality, enabling them to address potential issues early in the development process.
\end{abstract}
\maketitle

\section{Introduction}
Code smells were initially developed as a way to identify possible complications with the maintainability of software systems \cite{In1}. It is now being used as a metric of the design quality of software projects \cite{In2}, \cite{In3}, \cite{In4}. According to the findings of the researchers, code smells are connected to concerns in the code base relating to fault-proneness \cite{In6}, \cite{In7}, maintainability, and bug prevalence \cite{In3}, \cite{In5}. They have investigated the reasons that lead to the introduction of code smells, pointed the finger at several different contributors, such as developers having trouble meeting their deadlines \cite{In8} or not caring about the implications of the design decisions they made \cite{In1}.

Test code can also have code smells called "test smells". Van Deursen \textit{et al.}\cite{In9} said that test smells are caused by bad design decisions when making test cases. Similar to code smell, test smells make the affected test code harder to maintain and understand \cite{In10}. Recent studies have also shown that the quality of production code is also affected by the test code smells\cite{In11}.

Since test smells have a negative effect on the quality of production code, studying and detecting them is of utmost importance. To date, The majority of research on test smells so far has concentrated on statically typed languages such as Java, Scala, and Python \cite{In10}, \cite{In11}, \cite{In12}, \cite{In13}, \cite{In14}, \cite{In15}, \cite{In16}, \cite{In17}, \cite{In19}. In addition, an empirical study shows that the developers tend to be unaware of the smells present in their tests mainly due to the lack of efficient tools\cite{In13}. In recent years, C\#'s popularity in the field of software development has been steadily increasing\cite{In18}. To the best of our knowledge, no works investigate the existence and prevalence of test code smell in C\# code, and there are no tools that particularly aim at identifying test smells in this language.
{The implementation of a test smell detection tool specifically for C\# projects would have significant impacts, including improved test quality, enhanced maintainability, and reduced bug density. While existing static analysis toolsets (NDepend\cite{ndepend}, Roslyn-based analyzers\cite{roslyn}) in C\# offer valuable capabilities, they often lack dedicated support for detecting test smells. Obstacles to detecting test smell in C\# include language-specific syntax and semantics, integration with testing frameworks, framework-specific smells, and limited tooling and ecosystem support. By addressing these challenges and tailoring detection techniques to C\#, the proposed implementation would provide valuable support for C\# developers and testers, promoting better testing practices and software quality.}

In this paper, We intend to address these shortcomings by compiling a list of potential C\# test smells, a method for detecting them, a dataset that can act as a benchmark for future test smell detection, and an empirical study of their pervasiveness in C\# code. We began by conducting a small-scale mapping analysis to identify distinct test smells studied in the literature and picking test smells that can be regarded language-agnostic or have functionally equivalent counterparts in the C\# standard \textit{xUnit} framework. We found a total of 16 distinct test smells. All of these, 16 test smells were collected from articles focusing on other programming languages (for example \textit{Conditional Test Smell}, \textit{Empty Test}), although it is reasonable to suppose that C\# has its own unique test smells. 
Our investigation into identifying test smells involved scrutinizing the modification patterns in the test suites of various well-known repositories on open-source platforms like \textit{GitHub}. To achieve this, we searched for recurring changes that could indicate the presence of a test smell.
We evaluated 50 projects and identified 23 possible changes related to \textit{Assert} function in \textit{xUnit} testing framework and focused on increasing the level of specificity of the tests while also making the testing logic easier to understand. We identified this pattern using third-party software called NDepend\cite{ndepend}. This tool has advanced code evolution and code diff features that can distinguish between code changes such as method behavior change\cite{ndepend}. However, these patterns were already listed in the existing test smell literature. So we proceed our research with the already identified test smells.

To summarize, our contributions are as follows:\begin{itemize}
    \item[$\bullet$] A preliminary mapping research was carried out, and we developed a list of test smells that are relevant to C\#.
    \item[$\bullet$] We developed a tool called \textit{xNose} which can be used both as a Visual Studio extension or command line tool and able to detect test smells from C\# projects which usages \textit{xUnit} framework. 
    \item[$\bullet$] A prevalence study on test smells in \textit{xUnit}-based C\# projects conducted on 200 open-source projects.
\end{itemize}

\section{Related Works}
Test code, much like production code, is required to adhere to the best programming principles that have been developed \cite{R20}. Van Deursen \textit{et al.} established the notion of "test smells," which are code smells created by poor design decisions when creating test cases\cite{In9}. They identified a catalog of 11 test smells to go along with their introduction of the concept. Since that time, a number of researchers have added to the information contained in this catalog \cite{In14}, \cite{R21}, \cite{R22}, \cite{R23}. Even while the vast majority of studies have concentrated on Java test smells, researchers have also looked into test smells in other languages and domains. For instance, Bleser \textit{et al.} examined test smells in the Scala programming language \cite{In15}, \cite{R24}, whereas Peruma \textit{et al.} discovered several new test smells in unit tests for mobile applications \cite{R25}.

In addition, researchers \cite{In10}, \cite{In11}, \cite{In12}, \cite{In13}, \cite{R26} have investigated the negative consequences that test smells have on the development of software. Bavota \textit{et al.} carried out two empirical studies and came to the conclusion that test smells are prevalent in software systems and have a significant detrimental effect on the comprehensibility of test suites and production code \cite{In10}, \cite{In12}. Spadini \textit{et al.} investigated the connection between the presence of test smells and the likelihood of change and defects in test code, as well as the likelihood of defects in production code that has been tested. They identified that some test code smells are more prone to change than others, and the production code that is tested by smelly tests tends to be more prone to having defects \cite{In11}. Tufano \textit{et al.} \cite{In13} discovered that test smells are frequently present in a system when the corresponding test code is originally committed to the repository, and they have a tendency to remain there for a significant amount of time. Another group of researchers Virgínio \textit{et al.} explored correlations between test coverage and test smells and observed that test smells had an impact on code coverage \cite{R26}.

The research community has also shown interest in the concept of automating the detection of test smells. Van Rompaey \textit{et al.} developed a set of metrics that were described in terms of unit test principles \cite{R27}, and they contrasted the efficacy of their detection approaches with human review. Greiler Greiler \textit{et al.} investigated whether or not there was a correlation between the creation of a test fixture and the presence of smells that could be associated with the test.
They also built a static analysis tool \cite{In14} to identify fixture-related test smells and assessed it by finding test smells in three different industrial projects. This allowed them to determine whether or not the tool was effective.
Palomba \textit{et al.} devised an automated text-based approach to detect various sorts of test smells and found that it was more effective in detecting specific test smells than code metric-based strategies\cite{R28}. \textit{TSDETECT} was recently developed by Peruma \textit{et al.} as a tool that is capable of identifying 19 test smells in Java language based projects \cite{R25}, \cite{R29}.

In more recent times, researchers have started looking into approaches to aid testers with refactoring test smells. Lambiase \textit{et al.} introduced an IntelliJ-based plugin that utilizes the IntelliJ Platform's APIs to automatically detect and refactor test smells\cite{R30}. For automated detection of lines of code influenced by test smells and semi-automated refactoring for Java projects, Santana \textit{et al.} presented another tool that can be used in an IDE \cite{R31}. A technique for evaluating the overall quality of a test suite based on the presence or absence of certain "test smells" was created by Virgínio \textit{et al.} Theirs is the first tool to use code coverage and test smells together as a single metric for evaluating test quality \cite{R32}. In order to detect 18 different "test smells" in Python projects, Wang \textit{et al.} created an IntelliJ-based plugin \cite{In17}.


\section{Methodology}
This section contains information on how the authors collected the dataset for testing 'xNose' and the architecture of how this tool identifies test code smell.
\subsection{Data Collection}\label{sec:data-collection}
\par In order to create a benchmark dataset of test code written using the \textit{xUnit} package in C\#, we searched \textit{GitHub}. Since no existing dataset met our requirement which is labeled test code written in C\# , we utilized the advanced search option provided by GitHub. Our search query was: 

\begin{tcolorbox}[width=\linewidth-1em, colframe=black, colback=anti-flashwhite!30, boxsep=1mm, arc=1.5mm]
\textit{Github Query:} "stars: $\geq$ 100 language: C\# license: mit" and the results were sorted by "Most Stars". This resulted in over 4500 projects. 
\end{tcolorbox}
To build our dataset, we manually reviewed each project and selected the top 10 projects(sorted by 'stars') that contained test code written using the \textit{xUnit} testing framework in C\#. Although there are other testing packages available for this language, such as \textit{NUnit} and \textit{MSTest}, we considered \textit{xUnit} for our experiment. Therefore, we chose projects that exclusively used the \textit{xUnit} testing framework to ensure that our dataset is relevant and accurate for evaluating the performance of our tool. \par
Table \ref{tab:dataset} indicates the summary of our dataset. This dataset is also available in our tool repository for convenience. 
\begin{table*}[t]
    \caption{Test Dataset Collection}
    \label{tab:dataset}
    \centering
    \begin{tabular}{|c|c|c|c|}
    \hline
\textbf{Project} & \textbf{Test projects} & \textbf{Test class count} & \textbf{Test method count} \\ \hline
Aspnetboilerplate\cite{aspnetboilerplate} & 27 & 462 & 1818 \\ \hline
NLog\cite{nlog} & 4 & 322 & 4894 \\ \hline
C4Sharp\cite{c4sharp} & 1 & 1 & 5 \\ \hline
Ocelot\cite{ocelot} & 4 & 243 & 1108 \\ \hline
IdentityServer4.Admin\cite{skoruba} & 4 & 35 & 196 \\ \hline
Scrutor\cite{scrutor} & 1 & 48 & 923 \\ \hline
Refit\cite{refit} & 1 & 53 & 2395 \\ \hline
GraphQl-Platform.GreenDonut\cite{greendonut} & 1 & 15 & 119 \\ \hline
GraphQl-Platform.HotChocolate.Caching\cite{hotchocolatecaching} & 3 & 15 & 170 \\ \hline
GraphQl-Platform.HotChocolate.Core\cite{hotchocolatecore} & 22 & 1261 & 27183 \\ \hline
eShopOnWeb\cite{eshoponweb} & 4 & 27 & 50 \\ \hline
\textbf{Total} & 79 & 2720 & 39703\\ \hline
\end{tabular}
\end{table*}

\subsection{Selecting Test Smells}
The goal of our research is to create a tool that can detect test code smells in C\# test suites. First, we conducted a limited-scale mapping research of test smells to create a thorough list of test smells that have been discussed in the academic literature. The goal of a mapping study, as stated by Kitchenham \textit{et al.} \cite{M33}, is to take stock of what is already known about a certain subject.

Our search query was framed as follows\cite{In17}: \textit{"What test smells have been investigated in the literature thus far?"} To determine the most effective search terms, we piloted a search on two prominent digital libraries, IEEE and ACM. This process helped us to identify relevant keywords used in publications on test smells. Our query was limited to the titles and abstracts of publications to avoid erroneous matches. The final search string is provided below.

\begin{tcolorbox}[width=\linewidth-1em, colframe=black, colback=anti-flashwhite!30, boxsep=1mm, arc=1.5mm]
\textbf{\textit{Title:}}("test smell" OR "test smells") AND     \textbf{\textit{Abstract:}}"test smell" OR "test smells"
\end{tcolorbox}

In our search to locate relevant publications, we employed three of the most commonly used online paper search engines: ACM Digital Library, IEEE Xplore, and Scopus. To obtain the maximum number of related works, we looked up all relevant studies prior to 2023. This process yielded a bibliography of articles published between 2006 and 2022.

Our initial search yielded around 65 publications from the three digital libraries. We narrowed down the results by filtering out publications that did not meet our inclusion criteria. A summary of the inclusion and exclusion criteria used to shift through the retrieved literature is presented in Table \ref{tab:dataCollection}.
\begin{table}[h]
\caption{Inclusion and Exclusion Criteria\cite{In17}}
    \label{tab:dataCollection}
    \centering
    \setlength{\tabcolsep}{3pt}
    \normalsize
    \begin{tabular}{l}
         \hline
         \textbf{Inclusion Criteria} \\
         \hline
         1. Publications that implement software engineering \\ methodologies, approaches, and practices in test smell \\ detection and refactoring.\\
         2. Available in digital format.
         \\ \hline
         \textbf{Exclusion Criteria} \\ \hline
            1. Publications that are not written in English.\\
            2. Websites, leaflets, and grey literature.\\
            3. Published in 2023.\\
            4. Full-text is not available online.\\
            5. Tools not associated with peer-reviewed papers.\\
            6. Duplicated publications.\\
        \hline
    \end{tabular}
    \vspace{-3mm}
\end{table}

To ensure the credibility of our selected studies, each work was evaluated by three authors of this paper. This resulted in 35 different test smells encountered in Java, Scala, Android, and Python systems. Next, we considered the possibility of implementing each test smell for C\#. There were several reasons why some of the test smells could not be implemented:

The production code is required for the test smell detection to function properly. For instance, for identifying Lazy Test\cite{In9}, we need to know the production files and classes that correspond to it. Numerous recent studies have examined the topic of test-to-code traceability \cite{M38}, \cite{M39}, \cite{M40}, and a wide range of approaches have been proposed. Nonetheless, establishing a dependable 1-to-1 relationship between a production method and a test method within the context of static analysis presents a challenge \cite{M39}. As such, we have opted to defer addressing support for such test smells to future research.

Detecting test smells can only be done when the test is being executed. For instance, with the Test Run War \cite{In9}, it is essential to execute the test case, which cannot be done in a static analysis environment. Even after running, identifying such test smells is not straightforward, and we had to exclude them for practical purposes.

Finally, we selected 16 test smells for implementation. We listed them below:

\textbf{Assertion Roulette} arises when a test case contains numerous non-documented assertions. Having multiple assertion statements lacking a descriptive message can have adverse effects on the readability, comprehensibility, and maintainability of the code, as it becomes increasingly challenging to comprehend the reason for test failures \cite{In9}.

\textbf{Conditional Test Smell} occurs when test methods contain conditional logic such as \textit{if-else} statements or \textit{switch} cases. This type of test smell can lead to test methods that are difficult to understand, maintain and debug, as the different conditions and outcomes can create complex and convoluted test code\cite{In9}.

\textbf{Inappropriate Assertions} occurs when an inappropriate assertion is being used in the test code e.g., \textit{assertTrue} is used to check the equality of values instead of \textit{assertEquals}\cite{inppropriate-assertion}.

\textbf{Constructor Initialization} occurs when developers are unaware of the purpose of the \textit{IUseFixture} or \textit{IDisposable} interface \cite{xunitnet}, which contains the necessary preparations for executing test cases. Consequently, they define a constructor without using these interfaces for the test classes, which is not the best practice.

\textbf{Duplicate Assert} occurs when the same condition is tested in more than one place in a test case\cite{R25}.

\textbf{Empty Test} arises when a test case has zero executable statements. The code in such tests may have been commented out or they may have been generated for debugging purposes and forgotten.\cite{R25}.

\textbf{Eager Test} occurs when a test method invokes several methods of the production object. This smell results in difficulties in test comprehension and maintenance\cite{R25}.

\textbf{Ignored Test} results from ignoring test cases that can be suppressed from running. These ignored test cases increase code complexity and make comprehension more difficult, thus adding unnecessary overhead\cite{R25}.

\textbf{Lack of Cohesion} of Test Cases arises when test cases are grouped together in one test suite but lack cohesion. Coherence is a measure of the cohesiveness of a group or grouping, and it can be used to assess the degree to which roles and obligations within that group are shared. Issues with readability and maintenance can arise when a suite of tests isn't cohesive.\cite{R25}.

\textbf{Magic Number Test} happens when a test case contains assert statements that include numeric literals (i.e., magic numbers) as parameters instead of more descriptive constants or variables\cite{R25}.

\textbf{Obscure In-Line Setup} happens when a test case has too many setup steps, making it challenging to infer the purpose of the assertion in the test. In order to make the test more understandable, such preparation should be transferred to a distinct fixture or function.\cite{R25}.

\textbf{Redundant Assertion} refers to the situation where test methods contain assertion statements that are either always true or always false. Developers often introduce such assertions for debugging purposes but later forget to remove them. These assertions do not provide any useful information about the behavior being tested and can clutter the test code, making it harder to read and maintain. \cite{R25}.

\textbf{Redundant Print} occurs when a test contains a print statement. Such statements aren't needed in unit tests because they are usually run as part of an automated process that doesn't need much or any help from a human.\cite{R25}.

\textbf{Sleepy Test} occurs when developers pause a test case to simulate an external event before continuing. Since the processing time for a task varies on devices, explicitly putting a thread to sleep can produce unexpected outcomes\cite{R25}.

\textbf{Sensitive Equality} is a situation where a test method uses the toString() method to verify an object. This method calls the default toString() method of the object and compares the output with a specific string. Any changes to the implementation of toString() can result in a test failure. To avoid this, it is recommended to implement a custom method within the object to perform the comparison instead of relying on the toString() method\cite{R25}.

\textbf{Unknown Test} occurs when the test case does not contain any assertions. It is feasible to develop a test case that does not make use of any assertions; nevertheless, such a test would be more challenging to comprehend and analyze\cite{R25}.

Overall, we believe that our identification of these test smells can help improve the quality of C\# test code and ultimately lead to more reliable and effective software testing.

\subsection{\textit{xNose} Architecture}
\begin{figure*}[t]
    \centering
    \includegraphics[width=\linewidth]{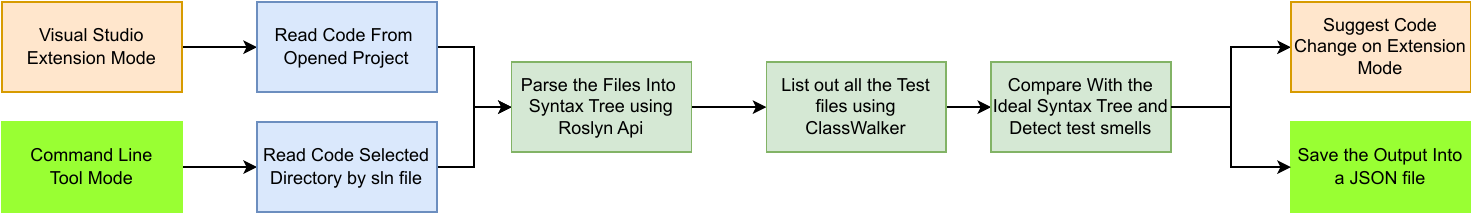}
    \caption{\textit{xNose} Architechiture}
    \label{fig:xNoseArc}
    \vspace{-3mm}
\end{figure*}

After curating the list of test smells (detailed in the previous section), our subsequent objective was to implement a tool capable of identifying these smells in actual C\# code. To achieve this, we created a tool named \textit{xNose}, which can currently detect 16 language-agnostic smells identified in existing literature, as described in the previous section. The tool can be executed from both a graphical user interface and a command line interface. The operating pipeline of \textit{xNose} is illustrated in Figure \ref{fig:xNoseArc}, which we will now explain in greater detail in this section. 
\textit{xNose} implemented as a visual studio\cite{vs} extension, a popular IDE for C\# developed by Microsoft. This supports two modes of operation: \textit{Command Line Interface Mode} and \textit{Visual Studio Extension Mode}. Internally, \textit{xNose} uses Roslyn APIs\cite{roslyn}(developed by Microsoft) to parse C\# source code and build syntactic and semantic code models for further analysis. After the project is started and the parser is configured, the tool leverages Roslyn and other relevant Visual Studio API to collect all test files and extract the Classes and the test Method declared into those test files.

Next, the tool extracts all the details about the classes and their methods and stores them in an intermediate class called \textbf{ClassVirtualizationVisitor} which allows \textit{xNose} to identify test smells. For example, for the \textit{Conditional Test Smell}, we use a custom smell visitor called \textbf{ConditionalTestSmell} to find all conditional, and then check if one of the provided arguments is a \textit{ConditionalExpression}. If there is a match, the Conditional Test smell is declared to be found.

To detect the test smells identified in existing literature, we implemented their detection using the same approach as described in their original papers, including the mentioned thresholds. For instance, we detect Obscure In-Line Setup in the same manner as Greiler \textit{et al.} \cite{In14}, by counting the number of local variables present in a test case, and flagging the test case as smelly if the count exceeds a threshold of 10. Similarly, we detect \textit{Lack of Cohesion of Test Cases} in the same way as Palomba \textit{et al.} \cite{R28}, by calculating pairwise cosine similarities between test cases in a test suit.

When the analysis is done, \textit{xNose} can show the detected test smells inside the Visual Studio IDE as suggestions in extension mode and save them to a JSON file for further analysis in Command Line Interface(CLI) mode.

This tool is designed in a way so that it can easily adapt to identify new test smells. If the user wants to collect new test smells then he/she just has to add a class that implements \textbf{ASmell} class, then implement the required \textit{HasSmell} method, and finally register this new class in \textbf{Program.cs} file. Then \textit{xNose} will detect this new test smell as well.


\section{Result Analysis}\label{sec:result-analysis}
An experimental assessment was carried out to determine the accuracy of \textit{xNose} in detecting test smells. Due to the unavailability of comprehensive datasets for all supported smells, a new validation set was constructed. We chose three projects (eShopOnWeb\cite{eshoponweb}, GraphQl-Platform.GreenDonut\cite{greendonut}, C4Sharp\cite{c4sharp}) from our actual dataset. The two authors manually labeled the test smells for each test case. These two authors have three to five years of experience in development in C\# and have experience in developing unit tests. 
After finalizing the validation set we ran \textit{xNose} on each project and calculated precision, recall, and F1 score for each test smell. We also calculated the weighted average of these three metrics for all test smells with the weights being the number of instances of each test
smell in the projects. The results of the conducted evaluation
are presented in table \ref{tab:validationResult}. 
\begin{table}[h]
   \caption{\textit{xNose} Result in Validation Set}
    \label{tab:validationResult}
    \centering
    \resizebox{\columnwidth}{!}{\begin{tabular}{|c|c|c|c|c|}
       \hline 
       \textbf{Test Smell} & \textbf{Instance} & \textbf{Precision} & \textbf{Recall} & \textbf{F1 Score} \\ \hline
       Lack Of Cohesion & 6 & 77.3\% & 84.2\% & 81\%\\ \hline
       Empty Test & 3 & 100\% & 100\% & 100\% \\ \hline
       Conditional Test Smell & 15 & 86\% & 100\% & 92.4\% \\ \hline
       Assertion Roulette & 17 & 100\% & 94.7\% & 97.3\% \\ \hline
       Unknown Test & 1 & 100\% & 100\% & 100\% \\ \hline
       Redundant Print & 2 & 100\% & 100\% & 100\% \\ \hline
       Sleepy Test & 2 & 100\% & 100\% & 100\% \\ \hline
       Ignored Test & 1 & 100\% & 100\% & 100\% \\ \hline
       Redundant Assertion & 3 & 100\% & 100\% & 100\% \\ \hline
       Duplicate Assert & 37 & 95.7\% & 94.6\% & 95\% \\ \hline
       Magic Number & 33 & 100\% & 87.8\% & 93.5\% \\ \hline
       Eager Test & 42 & 92.6\% & 95.3\% & 93.9\% \\ \hline
       Inappropriate Assertion & 2 & 100\% & 100\% & 100\% \\ \hline
       Sensitive Equality & 5 & 100\% & 80\% & 88.8\% \\ \hline
       Constructor Initialization & 1 & 100\% & 100\% & 100\% \\ \hline
       Obscure In-Line Setup & 5 & 100\% & 100\% & 100\% \\ \hline
       \textbf{Average}& - & 96.97\% & 96.03\% & 96.36\% \\ \hline
    \end{tabular}}
 \vspace{-3mm}
\end{table}

\begin{table*}[t]
\caption{xNose Report for Collected Projects}
    \label{tab:totalDataSet}
    \centering
    \resizebox{18.1cm}{!}{\begin{tabular}{|c|c|c|c|c|c|c|c|c|c|}
    \hline
        \textbf{Test Smell Type} & \textbf{Aspnetboilerplate} & \textbf{NLog} & \textbf{Ocelot} & \textbf{IdentityServer4.Admin} & \textbf{Scrutor} & \textbf{Refit} & \textbf{HotChocolate.Caching} & \textbf{HotChocolate.Core}  & \textbf{Total} \\ \hline
        Lack Of Cohesion & 31 & 24 & 6 & 0 & 0 & 2 & 0 & 14 & 77 \\ \hline
        Empty Test & 0 & 0 & 0 & 0 & 0 & 1 & 0 & 0 & 1 \\ \hline
        Conditional Test Smell & 4 & 463 & 2 & 0 & 14 & 0 & 0 & 99 & 582 \\ \hline
        Assertion Roulette & 0 & 489 & 0 & 0 & 218 & 665 & 90 & 139 & 1601 \\ \hline
        Unknown Test & 173 & 51 & 8 & 0 & 19 & 0 & 46 & 117 & 414 \\ \hline
        Redundant Print & 0 & 33 & 0 & 0 & 0 & 0 & 0 & 0 & 33 \\ \hline
        Sleepy Test & 17 & 151 & 2 & 0 & 0 & 0 & 0 & 0 & 170 \\ \hline
        Ignored Test & 3 & 248 & 2 & 0 & 0 & 2 & 0 & 35 & 290 \\ \hline
        Redundant Assertion & 0 & 43 & 2 & 0 & 0 & 0 & 0 & 0 & 45 \\ \hline
        Duplicate Assert & 670 & 541 & 171 & 96 & 318 & 225 & 12 & 511 & 2544 \\ \hline
        Magic Number & 48 & 989 & 4 & 0 & 294 & 202 & 84 & 790 & 2411 \\ \hline
        Eager Test & 300 & 523 & 13 & 5 & 18 & 97 & 170 & 218 & 1344 \\ \hline
        Sensitive Equality & 0 & 557 & 0 & 0 & 0 & 96 & 0 & 209 & 862 \\ \hline
        Constructor Initialization & 0 & 0 & 0 & 0 & 0 & 0 & 0 & 1 & 1 \\ \hline
        Obscure In-Line Setup & 1 & 37 & 0 & 21 & 0 & 7 & 0 & 0 & 66 \\ \hline
        Inappropriate Assertion & 10 & 158 & 0 & 0 & 0 & 66 & 42 & 0 & 276 \\ \hline
        \textbf{Total} & 1257 & 4343 & 210 & 122 & 881 & 1363 & 444 & 2133 & 10751 \\ \hline
    \end{tabular}}
\end{table*}

From the table \ref{tab:validationResult} it is clear that \textit{xNose} gained a high level of F1 scores for different test smells ranging from 81\% to 100\%. For the cases where our tool didn't achieve 100\% we manually investigated it.  In some instances, \textit{xNose} failed to identify the \textit{Magic Number Test Smell}. For example, \textit{Assert.Equal(\_testQuantity*3, result)} was tagged as a \textit{Magic Number Test Smell} by the human rater but our tool \textit{xNose} failed to identify this as the number was associated with another variable as a multiplication and our tool currently unable to identify if the number is associated with another variable. \textit{xNose} also made errors in detecting certain \textit{Conditional Test Logic} test smells. The presence of control statements such as if and for, regardless of their effect on the assertion, indicates the presence of \textit{Conditional Test Logic}. \textit{xNose} wrongly identifies some cases where the for statement is used only to assign a variable as \textit{Conditional Test Logic}. In measuring the cohesiveness of test cases in a test suite, \textit{xNose} uses cosine similarity, while human raters rely on subjective judgment. This difference in approach led to discrepancies between the output of \textit{xNose} and the opinions of human raters in multiple cases regarding the \textit{Lack of Cohesion smell}. 

Considering all the test cases together our tool \textit{xNose} achieves 96.97\% precision and 96.03\% recall value. 
\begin{table}[h]
\caption{Comparison of xNose with tsDetect and PyNose}
    \label{tab:comparision}
    \centering
    \begin{tabular}{|c|c|c|c|c|}
       \hline
       Tool & Language & Precision & Recall & F1\\ \hline
       tsDetect\cite{R29} & Java & 96.01\% & 97.11\% & 96.50\% \\ \hline
       PyNose\cite{In17} & Python & 94.00\% & 95.80\% & 94.90\% \\ \hline
       xNose & C\# & 96.97\% & 96.03\% & 96.36\% \\ \hline
    \end{tabular}
\end{table}
In table \ref{tab:comparision}, we present a comparison of the results obtained by our tool with those reported by tsDetect\cite{R29} and PyNose\cite{In17}, which are similar tools for Java and Python, respectively. The results show a similarity between the values obtained by our tool and those reported by the other tools. However, we intend to carry out a more comprehensive and direct comparison of these tools in the future to obtain a more thorough understanding of their similarities and differences. The details result for our dataset are given in table \ref{tab:totalDataSet}.

From the table \ref{tab:totalDataSet} it can be observed from the table that the top three test smells in terms of frequency are Duplicate Assert, Magic Number, and Assertion Roulette. These three test smells together account for more than half of the total test smells detected in the collected projects. It also indicates that the codebase of these projects may suffer from code duplication, lack of maintainability, and poor testing practices. This information can help the developers to focus on these specific test smells during code review and testing to improve the overall quality of their code.

However, it is important to remember that a low number of test smells does not necessarily indicate high code quality or that the code is bug-free. Therefore, developers should use these results as a starting point for further investigation and improvements in their testing and development practices.

\section{Prevalence of Test Smells}
After successfully validating \textit{xNose}, we conducted an empirical study on the prevalence of test smells on open-source C\# projects which contain test code using \textit{xUnit}. In this section, we present the details and the result of this study.
\subsection{Selecting projects}
The prevalence study was done to learn more about how test smells are distributed in C\# code and to enhance the subject diversity of the empirical studies on test smells that already exist. To ensure the study results are robust and do not depend on the results from Section-\ref{sec:result-analysis} we decided to include additional 200 GitHub projects in our dataset. To gather the dataset we used the same procedure described in Section - \ref{sec:data-collection} but this time our GitHub query was 
\begin{tcolorbox}[width=\linewidth-1em, colframe=black, colback=anti-flashwhite!30, boxsep=1mm, arc=1.5mm]
\textit{Github Query:} "topic: xUnit language: C\# license: mit" and the results were sorted by "Most Stars". This resulted in over 718 projects. 
\end{tcolorbox}
We took the top 200 projects from the response and drew our general conclusions from this updated dataset. The full list is available online\cite{emperical-dataset}. We studied the prevalence of the test smells and co-occurrence
of different test smells in individual test suites and discussed
the correlations between test smells.

\subsection{Methodology}
We ran \textit{xNose} on all of these projects. We only took into account test suites having at least one test case and test files with at least one test suite, dropping the results where not a single test suite was found. Test smells can appear at several degrees of granularity. Constructor initialization and lack of cohesion are examples of test smells that appear at the test suite level, while conditional test logic exists at the test case level. We analyzed the test smells using their appropriate granularity. A test suite is considered smelly if it contains at least one test case with a given smell. To obtain a more coarse-grained understanding of the test smell prevalence, we also calculated the distribution of test smells among projects. We looked at the most frequently encountered and least frequent test smells and the co-occurrence of various test smells in various test suites, and we addressed the connections between test smells.
\subsection{Results}
In this section, we  will discuss the results of our empirical study on the prevalence of test smells in open-source C\# repositories. These repositories contain a total of 394 test projects.

In total, at least one test case was found on 310 projects out of 394 projects (about $78.6\%$) in our dataset. From here on, all the percentages are calculated based on these 310 projects. In total, in these 310 projects, \textit{xNose} detected $7159$ test suits, and $149369$ test cases. More detailed statistics can be found in the table \ref{tab:prevalence-general}. It can be seen from the table that even mature projects vary greatly by the amount of testing within them. In our dataset, one test suite on average had 20.86 test cases. 
\begin{table}[h]
\caption{The summary of the amount of testing entities per project}
    \label{tab:prevalence-general}
    \centering
    \begin{tabular}{|c|c|c|}
       \hline
        & Test Suits & Test Cases \\ \hline
       Minimum & 1 & 1 \\ \hline
       Mean & 23.09 & 481.83 \\ \hline
       Maximum & 1137 & 9763 \\ \hline
    \end{tabular}
\end{table}

\subsubsection*{Distribution of Different Test Smells}
The smells of $16$ detected test smells are presented in table \ref{tab:smell-distribution}. From the table, it can be seen that the most common appeared test smell is the \textit{Duplicate Assert} that occurs in almost 43\% of projects. The other most common test smells are \textit{Assertion Roulette}, \textit{Magic Number}, and \textit{Eager Test} with an appearance of 32.90\%, 29.68\% and 28.07\% of projects respectively. From the study conducted on Python projects\cite{In17}, it can be seen that the type of most occurred test smells C\# projects are different from Python projects. The most appeared test smells in Python projects were \textit{Assertion Roulette} and \textit{Conditional Test Logic} with almost 90\% occurrence in the projects. 
\begin{table}[h]
\caption{The prevalence of different test smells among all projects and test suites}
    \label{tab:smell-distribution}
    \centering
    \begin{tabular}{|c|c|c|}
       \hline
        \textbf{Smell Type} & \textbf{Test Suits} & \textbf{Test Projects} \\ \hline
        Lack Of Cohesion & 13.17\% & 15.16\% \\ \hline
        Empty Test & 16.83\% & 4.84\% \\ \hline
        Conditional Test Smell & 7.08\% & 22.91\% \\ \hline
        Assertion Roulette & 33.36\% & 32.90\% \\ \hline
        Unknown Test & 23.12\% & 24.19\% \\ \hline
        Redundant Print & 0.56\% & 2.26\%  \\ \hline
        Sleepy Test & 0.87\% & 2.90\%  \\ \hline
        Ignored Test & 7.26\% & 9.81\%  \\ \hline
        Redundant Assertion & 4.25\% & 4.19\%  \\ \hline
        Duplicate Assert & 40.95\% & 42.77\% \\ \hline
        Magic Number & 27.93\% & 29.68\% \\ \hline
        Eager Test & 25.14\% & 28.07\%  \\ \hline
        Sensitive Equality &1.89\% & 3.03\% \\ \hline
        Constructor Initialization & 0.002\% & 0.01\% \\ \hline
        Obscure In-Line Setup & 2.19\% & 4.48\%  \\ \hline
        Inappropriate Assertion & 6.47\% & 12.26\% \\ \hline
        
    \end{tabular}
\end{table}
On the other hand, the least occurred test smells in C\# projects is the \textit{Constructor Initialization}. It occurred in only 0.01\% of the projects and spread out into 0.002\% of the test suites. The trends in the least occurred test smells are quite similar as the python projects \cite{In17} also have \textit{Constructor Initialization} as one of the rarest appeared test smells as well.

So the trends in the appearance of different test smells in Python and C\# language are not totally similar. Python and C\# programming languages do share syntactic and semantic similarities to some extent but there are a lot of dissimilarities between them as well. The dissimilarities between them may lead to different trends in test smell distribution. Further studies need to be done to reveal the causes behind those differences.

In conclusion, our findings reveal the prevalence of various test smells in C\# code. While certain test smells may be viewed as subjective, others present significant challenges in maintaining the codebase and interpreting test results during failures. Moving forward, we anticipate that \textit{xNose} can serve as a valuable tool for developers and researchers to address the propagation of test smells within their code repositories. By leveraging \textit{xNose}, developers can gain valuable insights and take proactive steps to improve code quality and test suite design, leading to more robust and maintainable software projects.

\subsubsection*{Co-occurrence of Test Smells}
\begin{figure}[h]
    \centering
    \includegraphics[scale=0.2]{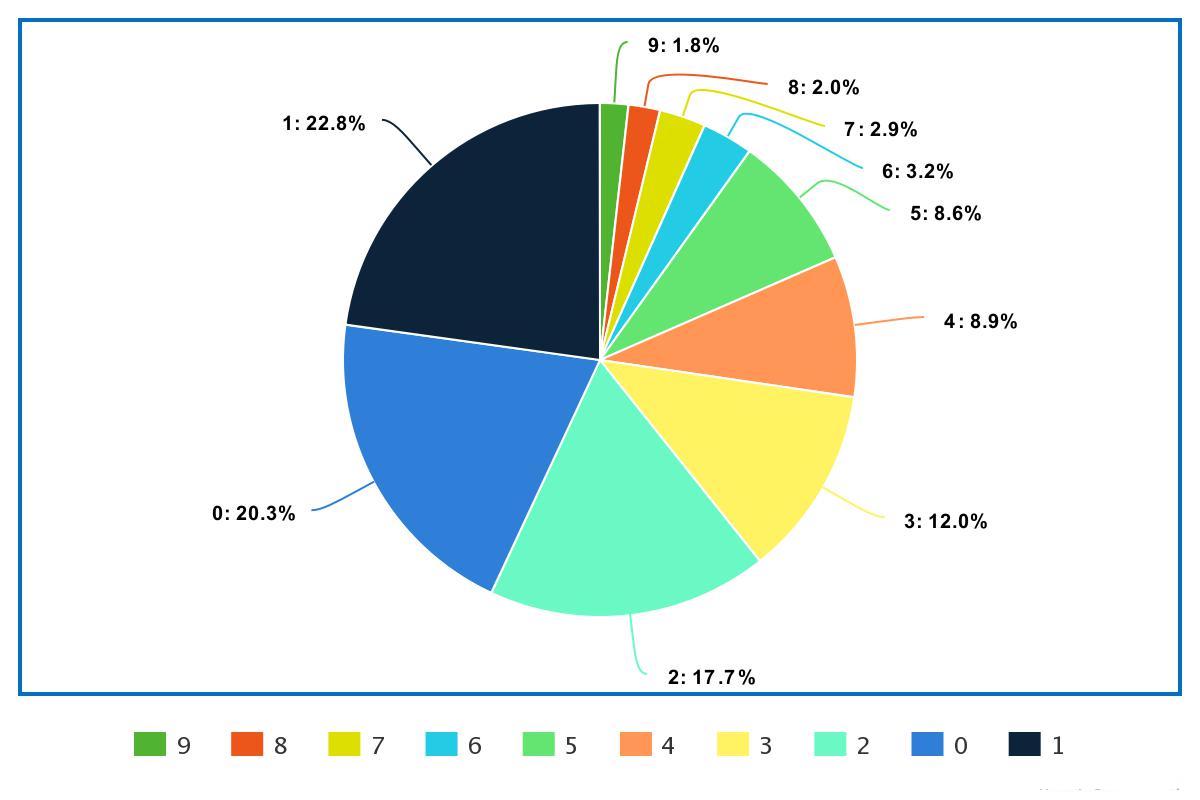}
    \caption{Co-occurrence of Different Test Smells}
    \label{fig:co-occur}
\end{figure}

In the previous section, we examined the prevalence of different test smells. However, the method used to analyze them treated each smell in isolation and might not fully capture the overall "smelliness" of the test codes. To gain a deeper insight, we also investigated how test smells co-occur. Figure \ref{fig:co-occur} depicts the occurrence of multiple test smells within a single test suite. Surprisingly, only $20.3\%$ of all test suites are completely free from any test smells. The majority, around $80\%$ of test suites, contain at least one test smell. Among them, $22.8\%$ have just one smell, $17.7\%$ exhibit two smells, and $12.0\%$ have three smells. As the number of co-occurring test smells increases, the proportion gradually declines.

Figure \ref{fig:co-occur} also provides valuable insights into the prevalence of test smells in C\# code. It is noteworthy that more than half of all test suites contain at least two different test smells, indicating a complex impact on code maintainability. We also conducted an in-depth analysis of specific pairs of test smells, calculating the percentage of test suites that exhibit both test smell X and test smell Y \cite{In17}. This highlights the significance of comprehending how these test smells interact and the potential impact they can have on code quality in C\#. Understanding the connections between test smells can provide valuable insights for improving test suite design, maintainability, and refactoring in software development.

Two pairs of test smells showed a complete connection. Firstly, when a test is \textit{Empty} (i.e., lacking any executable statements), it automatically becomes \textit{Unknown} (i.e., having no direct assertions). Additionally, other strongly connected pairs are associated with the widely observed test smell, \textit{Assertion Roulette}. For example, if a test suite contains a \textit{Duplicate Assert}, it is linked to \textit{Assertion Roulette} in $75.62\%$ of the cases. Similarly, $60.37\%$ of test suites with \textit{Redundant Assertion} also have \textit{Assertion Roulette}. This relationship is sensible, as \textit{redundant assertions} often imply the need for more meaningful assertions.

It is worth noting that the co-occurrence of these test smells in C\# code shares similar patterns with the co-occurrences observed in Python code \cite{In17}.  This may indicate that the co-occurrences of different test smells are somewhat language-independent. 

\section{Threats To Validity}
In order to maintain the integrity of our study and minimize the influence of chance factors, we took measures to prevent bias and reduce random noise. However, it is possible that our efforts to mitigate these risks were not entirely successful. In this section, we examine the potential threats to the validity of our research.

One possible limitation of our systematic mapping study of test smells is that we may have overlooked specific test smells that are relevant to C\#. Given the size and constant evolution of the C\# grammar, it is conceivable that we may have missed some test smell changes as a result. Additionally, we relied on \textit{NDepend}\cite{ndepend} for pattern detection. The limitations for detecting patterns of this tool apply to our study as well. Despite this, \textit{xNose} was designed to make it easy to incorporate new test smells in the future.

The result of our study relies on specific set of open-source C\# projects that we selected thus it might not be generalized to all projects. Although it is possible that \textit{xNose} may contain some errors that went unnoticed during its implementation, we took rigorous measures to minimize the risk. Specifically, we extensively tested the tool on synthetic data and manually evaluated it on real-world data. A potential threat to the validity of our study concerns the identification of certain test smells. In particular, some of the thresholds used to detect these smells were based on previous research and may not be optimal for C\#. Further investigation is needed to address this issue.

\section{Conclusion and Future Work}
Test smells are common in widely used programming languages like Java and Python, and they have a negative impact not only on the quality of the code used for testing but also on the code used in production. Additionally, test smells have a negative impact on the quality of test code, but they also have a negative impact on the quality of production code\cite{In11}.

In this research, we presented \textit{xNose}, the first test code smell detection tool for C\# language. This tool is capable of identifying 16 test code smells that were adapted from test smells for other programming languages that were described in the literature. Experiments on 6 real-world open-source projects consisting of 43 test classes and almost 200 test methods showed that \textit{xNose} achieved 96.97\% precision and recall value of 96.03\% in test smell code detection. The empirical study shows that the test smells are prevalent in C\# projects, with $80\%$ of the test suits have atleast one test smell in them and the most frequently detected code smells are \textit{Duplicate Assert}, \textit{Magic Number}, and \textit{Assertion Roulette}. We also believe the output dataset for this tool can act as a benchmark for further studies based on C\# test smell detection. 

Future research directions for this work includes: \begin{itemize}
    \item[$\bullet$] Discover C\# dependent test smell.
    \item[$\bullet$] Make a larger manually labeled dataset to conduct a thorough comparison of \textit{xNose} with other tools such as \textit{PYNOSE} and \textit{TSDETECT}.
    \item[$\bullet$] Detailed comparison of test smell co-occurrences across multiple programming languages.
\end{itemize}

\textit{xNose} is publicly available for research here:\\ \href{https://github.com/Partha-SUST16/xNose}{https://github.com/Partha-SUST16/xNose}, the results generated by our tool is also available there under the \href{https://github.com/Partha-SUST16/xNose/tree/main/results}{\textit{result}} folder.

\bibliographystyle{ACM-Reference-Format}
\bibliography{references}
\end{document}